\documentclass[]{fairmeta}
\usepackage{graphicx}
\usepackage{booktabs}
\usepackage{amsmath}
\usepackage{amssymb}
\usepackage{multirow}
\usepackage{bm}
\usepackage{bbding}
\usepackage{algorithmicx}
\usepackage{algpseudocode}
\usepackage{algorithm}
\usepackage{wrapfig}
\usepackage{tabularx}
\usepackage{textcomp}
\usepackage{stfloats}
\usepackage{url}
\usepackage{verbatim}
\usepackage{titlesec}
\usepackage{adjustbox}
\usepackage{pifont}
\usepackage{tikz}
\usepackage{comment}
\usepackage{colortbl}
\usepackage{xcolor}
\usepackage{xspace}

\usepackage{subfig} 

\usepackage{marvosym}
\usepackage{hyperref}

\RequirePackage{xspace}
\makeatletter
\makeatletter
\@ifundefined{sf@counterlist}{%
  \newcommand{\sf@counterlist}{}% 
}{}
\makeatother
\DeclareRobustCommand\onedot{\futurelet\@let@token\@onedot}
\def\@onedot{\ifx\@let@token.\else.\null\fi\xspace}

\makeatother

\definecolor{adptorange}{RGB}{248, 205, 172}
\definecolor{cmpblue}{RGB}{189, 215, 238}
\definecolor{cmpblue}{RGB}{189, 215, 238}

\definecolor{our_red}{RGB}{232,157,160}
\definecolor{our_blue}{RGB}{136,206,230}
\definecolor{our_orange}{RGB}{246,200,168}
\definecolor{our_green}{RGB}{178,211,164}

\definecolor{attn_code0}{RGB}{247,215,200}
\definecolor{attn_code1}{RGB}{238,169,139}
\definecolor{mlp_code0}{RGB}{204,201,221}
\definecolor{mlp_code1}{RGB}{102,95,153}

\definecolor{token_blue}{RGB}{84, 120, 140}
\usepackage{pifont}       % \ding{xx}
\usepackage{bbding}       % \Checkmark,\XSolid,... (需要和pifont宏包共同使用)
\usepackage{fontawesome}
\usepackage{xspace}

\usepackage{float}

\newlength\savewidth

\newcolumntype{x}[1]{>{\centering\arraybackslash}p{#1pt}}
\newcolumntype{y}[1]{>{\raggedright\arraybackslash}p{#1pt}}
\newcolumntype{z}[1]{>{\raggedleft\arraybackslash}p{#1pt}}

\renewcommand{\paragraph}[1]{\vspace{1mm}\noindent\textbf{#1}}
\usepackage{colortbl}
\usepackage{xcolor}
\usepackage{wrapfig}

\renewcommand{\paragraph}[1]{\vspace{1.25mm}\noindent\textbf{#1}}

\usepackage{algorithm}
\usepackage{listings}

\definecolor{codeblue}{rgb}{0.25, 0.5, 0.5}
\definecolor{codekw}{rgb}{0.35, 0.35, 0.75}
\lstdefinestyle{Pytorch}{
    language = Python,
    backgroundcolor = \color{white},
    basicstyle = \fontsize{9pt}{8pt}\selectfont\ttfamily\bfseries,
    columns = fullflexible,
    aboveskip=1pt,
    belowskip=1pt,
    breaklines = true,
    captionpos = b,
    commentstyle = \color{codeblue},
    keywordstyle = \color{codekw},
}

\definecolor{green}{HTML}{009000}
\definecolor{red}{HTML}{ea4335}

\title{PLUS: Plug-and-Play Enhanced Liver Lesion Diagnosis Model on Non-Contrast CT Scans}

\author[1, 2, 3]{Jiacheng Hao}
\author[2, 3(\textrm{\Letter})
]{Xiaoming Zhang}
\author[ 2,3]{Wei Liu}
\author[ 4]{Xiaoli Yin}
\author[ 2,3]{Yuan Gao}
\author[ 4]{Chunli Li}
\author[ 2]{Ling Zhang}
\author[ 2]{Le Lu}
\author[ 4]{Yu Shi}
\author[ 5]{Xu Han}
\author[ 2,3]{Ke Yan}

\affiliation[1]{School of Biomedical Engineering, Tsinghua University\\}
\affiliation[2]{DAMO Academy, Alibaba Group\\}
\affiliation[3]{Hupan Lab\\}
\affiliation[4]{Department of Radiology, Shengjing Hospital of China Medical University\\}
\affiliation[5]{Department of Hepatobiliary and Pancreatic Surgery, First Affiliated Hospital of
Zhejiang University\\}

\contribution[\textrm{\Letter}]{Corresponding Author: \email{zxiaoming360@gmail.com}}

\abstract{
Focal liver lesions (FLL) are common clinical findings during physical examination. Early diagnosis and intervention of liver malignancies are crucial to improving patient survival. Although the current 3D segmentation paradigm can accurately detect lesions, it faces limitations in distinguishing between malignant and benign liver lesions, primarily due to its inability to differentiate subtle variations between different lesions. Furthermore, existing methods predominantly rely on specialized imaging modalities such as multi-phase contrast-enhanced CT and magnetic resonance imaging, whereas non-contrast CT (NCCT) is more prevalent in routine abdominal imaging. To address these limitations, we propose PLUS, a plug-and-play framework that enhances FLL analysis on NCCT images for arbitrary 3D segmentation models. In extensive experiments involving 8,651 patients, PLUS demonstrated a significant improvement with existing methods, improving the lesion-level F1 score by 5.66\%, the malignant patient-level F1 score by 6.26\%, and the benign patient-level F1 score by 4.03\%. Our results demonstrate the potential of PLUS to improve malignant FLL screening using widely available NCCT imaging substantially. Code is available at \href{https://github.com/alibaba-damo-academy/plug-and-play-diagnosis}{https://github.com/alibaba-damo-academy/plug-and-play-diagnosis}.
}

\date{\today}

\begin{document}

\thispagestyle{firstheader}
\maketitle
\pagestyle{empty}

\section{Introduction} \label{sec:introduction}
\noindent 
The incidental detection of focal liver lesions (FLLs) during routine physical examination has become increasingly prevalent~\citep{marrero2014acg}. However, accurate differentiation between malignant and benign lesions remains a significant challenge for radiologists due to the subtle heterogeneity present in images. Benign lesions are usually harmless or require regular monitoring, whereas malignant tumors require immediate intervention as delayed diagnosis can lead to rapid disease progression and poor outcomes~\citep{bilic2023liver,marrero2014acg}. In this scenario, early detection and differentiation of the malignant lesions plays a crucial role in patient survival, with five-year survival rates exceeding 70\% for early-stage diagnoses, underscoring the critical importance of effective screening programs, particularly for high-risk populations with chronic liver conditions~\citep{de2022baveno}.

Contrast-enhanced CT (CECT) and magnetic resonance imaging (MRI) are primary modalities for FLL diagnosis due to their high sensitivity~\citep{huang2024lidia}. However, both face significant limitations in the context of large-scale opportunistic screening. CECT requires contrast agents that carry risks of nephrotoxicity and allergic reactions, restrict follow-up examination frequency, and substantially increase healthcare costs. MRI, a radiation-free imaging modality, is unsuitable for large-scale screening due to limited availability, lengthy protocols, and device contraindications. Meanwhile, non-contrast CT (NCCT) presents advantages for large-scale malignancy screening~\citep{cao2023large,yao2022effective}, combining widespread availability, cost-effectiveness, rapid acquisition, and absence of contrast-related risks, making it ideal for broad FLL screening.

Despite NCCT's advantages, its fundamental reliance on tissue density differences makes it challenging to detect FLLs, especially early-stage or subtle lesions with minimal distinction against surrounding tissues, inherently hindering detection and complicating classification due to subtle variations among tumor types. Chronic liver conditions also further complicate tumor identification, e.g., fatty liver, cirrhosis, and biliary dilation. While radiologists' expertise is crucial, the subtle imaging characteristics make diagnosis highly experience-dependent and prone to errors, highlighting the need for computer-aided methods. 

Recently, deep learning methods have shown promise in enhancing automatic FLL detection by accurately capturing the lesion patterns on CT images~\citep{wei2024focal}. 
%However, current deep learning-based approaches to automatic FLL analysis still face significant limitations.
Although well-configured CNN-based architectures remain competitive for general medical image segmentation~\citep{isensee2024nnu,wang2022cycmis,wang2022few}, FLL classification presents unique challenges beyond detection. Despite various architectural innovations including Transformers~\citep{Hatamizadeh2021UNETRTF}, and Mamba-based designs~\citep{ruan2024vmunet}, the accurate subtype classification remains a key challenge. MaskFormer-based approaches~\citep{Cheng2021MaskFormer,Cheng2022Mask2Former}, despite their promising results through dense cross-attention, lack explicit mechanisms to model crucial anatomical relationships between lesions and liver tissue. Existing classification methods~\citep{Tang2020E2Net,Zhang2021NC,wang2023cascaded,zhang2023spectral} typically focus on isolated lesion-level or patient-level predictions, overlooking the complex interplay between global liver morphology and local lesion characteristics. While recent approaches~\citep{ren2024hif,kareem2024improving3dmedicalimage,Tragakis_2024,Yan2023Plan} attempted global-local feature modeling, they failed to capture clinically relevant and multi-scale liver pathology information. Drawing insights from multi-organ cohesion analysis~\citep{li2024improved}, we propose a classification-focused framework that emphasizes feature discrimination within pre-detected regions of interest (ROIs), addressing the limitations of detection-based approaches.

To address these limitations, we propose a \textbf{PLU}g-and-play enhanced liver lesion diagnosi\textbf{S} model (PLUS), a framework compatible with arbitrary 3D segmentation models for improved FLL analysis on NCCT. PLUS introduces three key components: (1) a hierarchical dual attention (HDA) mechanism facilitating bidirectional exchange between global liver and local lesion features; (2) a graph-based prior reasoning (GPR) module selectively leveraging prior pretrained segmentation knowledge for FLL type classification; and (3) a combined optimization strategy for lesion-level detection with patient-level diagnosis.

We curated the largest NCCT FLL dataset containing 28,853 annotated lesions from 8,651 patients and healthy subjects. PLUS achieves F1 scores of 65.11\% for lesion-level detection, 90.12\% for malignant, and 73.10\% for benign patient-level diagnosis, surpassing other methods. Through extensive ablations and visualizations, we demonstrate PLUS's effectiveness and potential in enhancing large-scale FLL screening and differentiation via widely available NCCT.

\section{Methodology}
\noindent
\subsection{Problem Definition}
Considering a 3D NCCT volume $I \in \mathbb{R}^{H \times W \times D}$, the PLUS framework builds on an arbitrary pre-trained segmentation model $\phi(\cdot)$ that processes the input volume and generates segmentation masks $M = \phi_{\text{seg}}(I) \in \{0,1\}^{H \times W \times D}$ and preliminary lesion classification logits $P = \phi_{\text{cls}}(I) \in \mathbb{R}^C$, where $C$ denotes the number of lesion categories. The segmentation mask $M$ inherently contains two anatomical regions: the liver region $M_{H}$ and the set of lesion instances $\mathcal{S} = \{(M_{s_i}, P_i)\}_{i=1}^N$, which serve as spatial attention guides for subsequent feature extraction, where $N$ is the number of detected lesions, the objective of the plug-in framework $f_{\theta}(\cdot)$ is to leverage these segmentation-derived priors and enhance the final classification prediction through $f_{\theta}(I, M_{H}, \mathcal{S})$.

\begin{figure}[t!]
	\centering
	\includegraphics[width=1\textwidth]{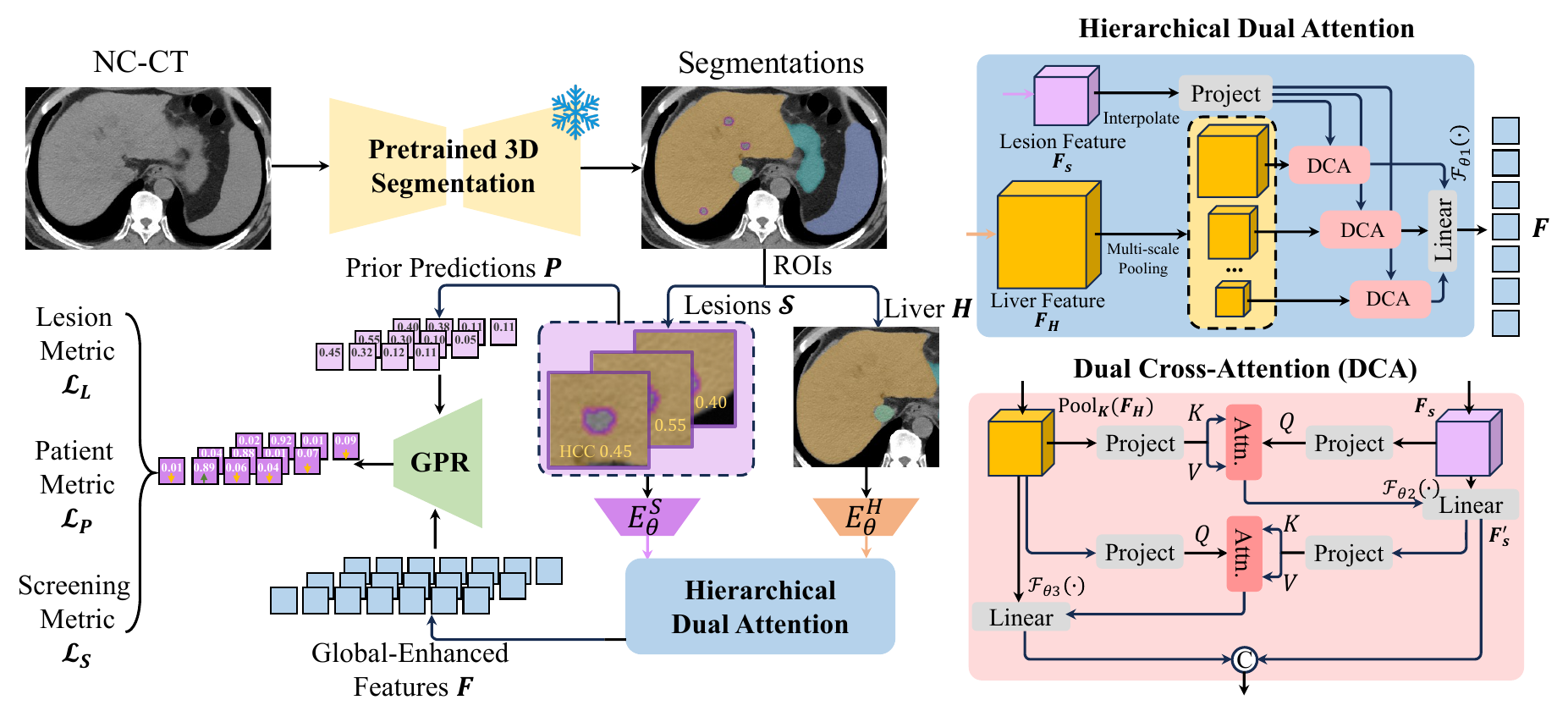}
	\caption{Illustration of the overall pipeline of PLUS.}
	\label{fig:framework}
\end{figure}
\begin{figure}[ht!]
	\centering
	\includegraphics[width=0.8\textwidth]{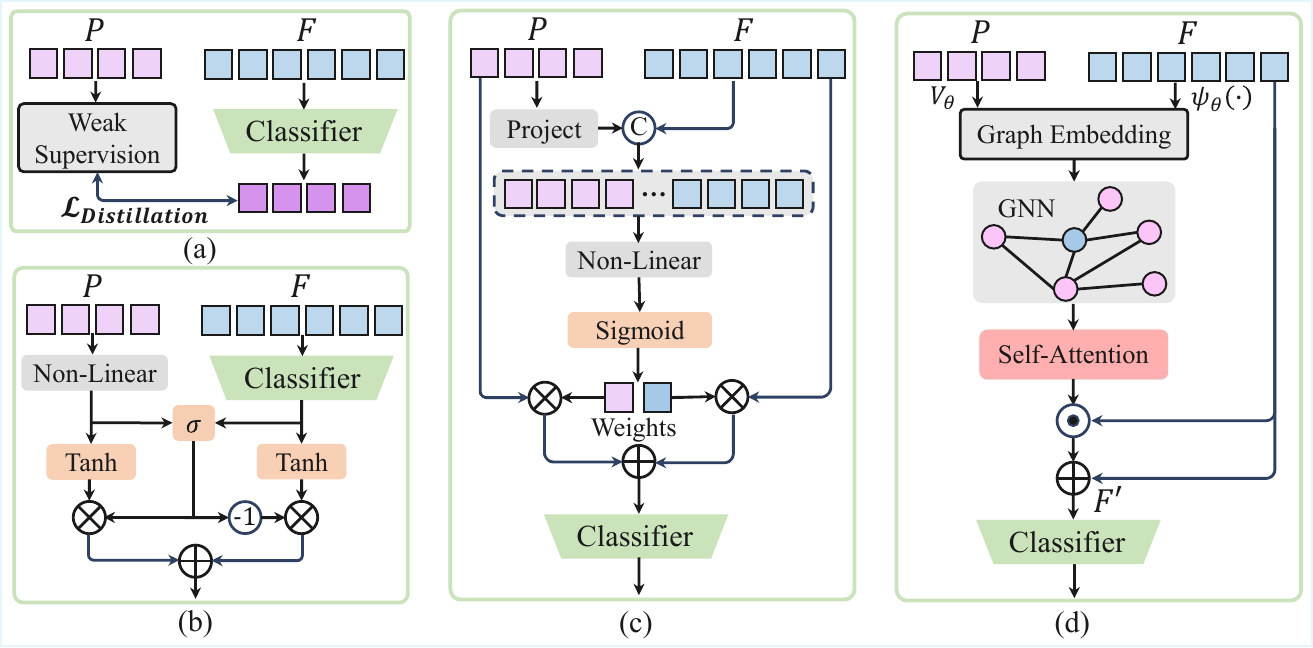}
	\caption{Comparison of classifier enhancement strategies: (a) Knowledge Distillation, (b) Gated Fusion, (c) Weighted Fusion, (d) Graph-Based Prior Reasoning.}
	\label{fig:GPR}
\end{figure}
\subsection{Plug-and-Play Enhanced Liver Lesion Diagnosis}
For the input NCCT volume $I$ and its segmentation masks of liver $M_{\text{H}}$ and pre-segmented lesion instances $\mathcal{S} = \{(M_{s_i}, P_i)\}_{i=1}^N$ from model $\phi(\cdot)$, the framework extracts region-specific features using two encoders: $F_H = E_{\theta}^{H}(I \odot M_H)$ for liver tissue and $F_{s_i} = E_{\theta}^{S}(I \odot M_{s_i})$ for each lesion instance $i$, see Fig.~\ref{fig:framework}. The resulting feature tensors $F_{H}$ and $\{F_{s_i}\}_{i=1}^N$ encode contextual information from the liver and detailed characteristics of individual lesions, respectively.

\noindent
\textbf{Hierarchical Dual Attention.} 
The initial representations lack cross-scale feature interaction between anatomical context and local lesion patterns. To enable effective multi-scale feature interaction, we propose HDA, a module enabling bidirectional intra-/inter-scale semantic fusion with liver and lesions. 

Given lesion features $F_s$ and liver features $F_H$, the HDA mechanism can be formulated as concatenated features
    $\textbf{F} = \mathcal{F}_{\theta1}([{\text{DCA}(\text{Pool}_i(F_H);F_S)}]_{i=1}^K)$, where the dual cross-attention (DCA) combines bidirectional attention flows:
\begin{equation}
   F_{x}'=\mathcal{F}_{\theta2}(\mathcal{A}(F_x,F_y)),\quad \text{DCA}(F_{x}, F_{y}) = \big[F_{x}'; \mathcal{F}_{\theta3}(\mathcal{A}(F_{y}, F_{x}'))\big],
\end{equation} 
with attention operation $\mathcal{A}(F_a, F_b) = \text{Attn.}(\mathcal{T}_{Q}(F_a), \mathcal{T}_{K}(F_b), \mathcal{T}_{V}(F_b))$, $\mathcal{T}_{\{Q,K,V\}}$ are query/key/value projection operations, $\text{Pool}_i(\cdot) $ represents multi-scale pooling at the $ i $-th scale, $\mathcal{F}_{\theta}$ are learnable linear projections, and the standard attention operation is defined as $\text{Attn.}(\mathbf{Q}, \mathbf{K}, \mathbf{V}) = \text{softmax}\left(\frac{\mathbf{Q}\mathbf{K}^\top}{\sqrt{d}}\right)\mathbf{V}$.

The design of HDA is motivated by two key clinical observations in liver lesion diagnosis. First, radiologists zoom in and out the image when examining lesions, as different lesions exhibit distinctive patterns on various scales. HDA mimics this diagnostic process through efficient multi-scale feature interaction with progressive pooling operations. Furthermore, the interpretation of the lesion depends heavily on the surrounding liver conditions, where bidirectional interaction is crucial. For example, a suspicious lesion of a cirrhosis patient tends to be a hepatocellular carcinoma, while a low-density region on the liver of a steatosis patient is probably a false positive tumor. 
\begin{algorithm}[t!]
	\caption{Graph-based Prior-Aware Reasoning}
	\label{alg:GPR}
	\begin{algorithmic}[1]
		\Require Features $\mathbf{F} \in \mathbb{R}^{B \times D}$,  Prior Prediction $\mathbf{P} \in \mathbb{R}^{B \times C}$, Prototypes $\mathbf{V}_\theta \in \mathbb{R}^{C \times D}$
		\Ensure Prior-aware Features $\mathbf{F}' \in \mathbb{R}^{B \times D}$
		
		\State $\mathbf{F}_{\psi} = \psi_{\theta}(\mathbf{F}), \mathbf{V}_w = \mathbf{P}_{\psi} \times \mathbf{V}_\theta$ \Comment{Feature projection and weighted prototypes}
		\State $\mathbf{G} = [\mathbf{F}_\phi; \mathbf{V}_\theta]$ \Comment{Node set construction: $B$ original nodes $+$ $B$ prototypes nodes}
		\State $\mathbf{Q}, \mathbf{K}, \mathbf{V} = \mathcal{W}_q(\mathbf{G}), \mathcal{W}_k(\mathbf{G}), \mathcal{W}_v(\mathbf{G})$; $\alpha_{ij} = \text{softmax}(\frac{\mathbf{q}_i^T\mathbf{k}_j}{\sqrt{d}}), \, \mathbf{m}_i = \sum_{j=1}^{2B} \alpha_{ij}\mathbf{v}_j$ 
		\State \Return $\mathbf{F}’=\mathbf{F}+\mathcal{F}_{\theta}([\mathbf{F}_{\phi} \parallel \mathbf{m}_{1:B}])\odot \mathbf{V}_w$ \Comment{Updated feature of central node}
        
	\end{algorithmic}
\end{algorithm}

\noindent\textbf{Graph-based Prior-aware Reasoning.} 
We propose a graph-based prior-aware reasoning (GPR) module to refine potentially inaccurate prior lesion predictions through graph neural networks~\citep{kipf2016semi}. As outlined in Algorithm~\ref{alg:GPR} and (Fig.~\ref{fig:GPR}\textcolor{red}{(d)}), GPR constructs a homogeneous graph $\textbf{G}$ where enhanced features $\textbf{F}$ serve as center nodes, connected with class-specific prototypes $V_\theta$ as surrounding nodes. Instead of directly using the initial prior prediction $\textit{P}$, which may be unreliable~\citep{meng2021graph}, we leverage it as a conditional message to guide prototype-based reasoning. Through message passing and self-attention~\citep{vaswani2017attention} between features and prototypes, the graph-based structure automatically learns to enhance relevant prior knowledge while suppressing unreliable predictions, enabling more robust feature enhancement compared to conventional fusion methods.

\noindent
\textbf{Training Paradigm.} 
Clinically, the diagnosis of FLL through medical imaging follows a systematic process: radiologists first analyze individual lesion characteristics, and then compile these findings into patient-level diagnoses. Both lesion-level and patient-level analyses are crucial: the former provides detailed lesion-level insights while the latter guides overall treatment planning. This is particularly challenging in cases with multiple lesions, where effective aggregation of lesion-level information becomes essential for accurate patient-level assessment. Inspired by this clinical workflow, we propose a comprehensive optimization strategy that aligns with radiologists' decision-making process. 

The lesion level loss $\mathcal{L}_L = -\frac{1}{N}\sum_{i}^{N}\sum_{c}^{C} \mathbf{z_c}(1-p_{i,c})^\gamma y_{i,c}\log p_{i,c}$ addresses class imbalance through lesion-specific weights $z_c$, and focal parameter $\gamma$, with $N$ lesions across $C$ categories. The overall lesion-level loss is the average of the individual losses for each patient's lesions. Taking malignancy prediction as example, the patient-level loss $\mathcal{L}_P = -\frac{1}{M}\sum_{j}^{M} [Y_j\log( \max_{i\in \mathcal{S}_j}p^{j}_{i,malig})+(1-Y_j)\log(1- \max_{i\in \mathcal{S}_j}p^{j}_{i,malig})]$ aggregates malignancy probabilities over $M$ patients, prioritizing clinically critical findings. Screening loss $\mathcal{L}_S = -\frac{1}{M}\sum_j^M [y_j\log q_j + (1-y_j)\log(1-q_j)]$ uses $q_j = \max_{i\in \mathcal{S}_j} (p_{i,malig} \parallel p_{i,beni.})$ to detect whether each subject has any tumor. 
The overall loss function integrates three hierarchically structured components $\mathcal{L}_{\text{total}} = \alpha \mathcal{L}_{L} + \beta \mathcal{L}_P + (1-\alpha-\beta) \mathcal{L}_S$. This tripartite design validates PLUS through the diagnostic cascade: $\text{screening} \rightarrow \text{lesion analysis} \rightarrow \text{diagnosis}$, ensuring consistency across clinical workflow stages.

% \usepackage{multirow}
\begin{table}[ht!]
    \centering
    \caption{Performance comparison on test set. ($\bullet$: malignant, $\circ$: benign.)}
    \resizebox{\linewidth}{!}{
\begin{tabular}{c|ccc|cccccc|cc}
\hline
\multirow{2}{*}{Method}                                       & \multicolumn{3}{c|}{Lesion-level}                & \multicolumn{6}{c|}{Patient-level Diagnosis}                                                                   & \multicolumn{2}{c}{Screening}   \\ \cline{2-12} 
                                                              & F1             & Prec.          & Recall         & F1$\bullet$    & Prec.$\bullet$ & Recall$\bullet$ & F1$\circ$      & Prec.$\circ$   & Recall$\circ$  & F1             & Acc.           \\ \hline
nnUNet~\cite{Isensee2021nnu}            & 56.19          & 46.43          & \textbf{71.16} & 80.91          & 74.93          & \textbf{87.92}  & 70.34          & 75.27          & 66.02          & 87.45          & 88.02          \\
+PLUS                                                         & \textbf{62.97} & \textbf{57.62} & 69.42          & \textbf{83.74} & \textbf{81.56} & 86.04           & \textbf{72.68} & \textbf{77.53} & \textbf{68.41} & \textbf{87.83} & \textbf{88.45} \\ \hline
Mask2Former~\cite{Cheng2022Mask2Former} & 58.19          & 49.56          & \textbf{70.48} & 82.80          & 80.32          & \textbf{85.45}  & 67.51          & 72.02          & 63.54          & 87.78          & 88.14          \\
+PLUS                                                         & \textbf{63.46} & \textbf{58.85} & 68.87          & \textbf{86.02} & \textbf{89.32} & 82.97           & \textbf{72.18} & \textbf{76.54} & \textbf{68.30} & \textbf{87.96} & \textbf{88.84} \\ \hline
PLAN~\cite{Yan2023Plan}                 & 59.45          & 48.27          & \textbf{77.39} & 83.86          & 75.38          & \textbf{94.49}  & 69.07          & 73.63          & 65.05          & 88.89          & 89.00          \\
+PLUS(distill.)~\cite{gou2021knowledge} & 61.08          & 58.48          & 63.94          & 88.47          & 87.58          & 89.38           & 71.39          & 79.67          & 64.67          & 88.39          & 89.08          \\
+PLUS(gated)~\cite{arevalo2017gate}     & 57.92          & 51.47          & 66.24          & 88.53          & 86.49          & 90.68           & 66.39          & \textbf{84.93} & 54.50          & 88.82          & 88.96          \\
+PLUS(weighted)~\cite{yin2018tensor}    & 64.47          & 56.38          & 75.26          & 87.84          & 86.12          & 89.65           & 64.41          & 76.75          & 55.50          & \textbf{89.04} & 89.42          \\
+PLUS(ours)                                                   & \textbf{65.11} & \textbf{60.28} & 74.56          & \textbf{90.12} & \textbf{88.56} & 91.74           & \textbf{73.10} & 76.60          & \textbf{69.90} & 88.97          & \textbf{89.73} \\ \hline
    \end{tabular}}
    \label{tab:performance}%
\end{table}%

\begin{table}[ht!]
    \centering
    \caption{Ablation study on proposed components. ($\bullet$: malignant, $\circ$: benign.)}
    \begin{tabular}{c|ccc|cccccc}
        \hline
        \multirow{2}{*}{Method} & \multicolumn{3}{c|}{Lesion-level} & \multicolumn{6}{c}{Patient-level Diagnosis} \\
        \cline{2-10}
                                 & F1            & Prec.         & Recall        & F1$\bullet$       & Prec.$\bullet$       & Recall$\bullet$      & F1$\circ$           & Prec.$\circ$       & Recall$\circ$       \\
        \hline
        PLAN                     & 59.45         & 48.27         & \textbf{77.39} & 83.86        & 75.38        & \textbf{94.49} & 69.07        & 73.63        & 65.05        \\
        +HDA                     & 64.67         & 58.72         & 71.96         & 89.25        & 87.53        & 91.05        & 71.56        & 75.33        & 68.16        \\
        +GPR                     & \textbf{65.87} & \textbf{60.51} & 72.23         & 89.54        & 88.24        & 90.89        & {71.84} & {76.12} & 68.02        \\
        +Comb. Loss              & {65.11} & 60.28 & {74.56} & \textbf{90.12} & {\textbf{88.56}} & {91.74} & {\textbf{73.10}} & {\textbf{76.60}} & {\textbf{69.90}} \\
        \hline

    \end{tabular}%
    \label{tab:performance_comparison}%
\end{table}%
\begin{figure}[!t]
    \centering
    \begin{subfigure}{0.25\textwidth}
        \centering
        \includegraphics[width=\linewidth]{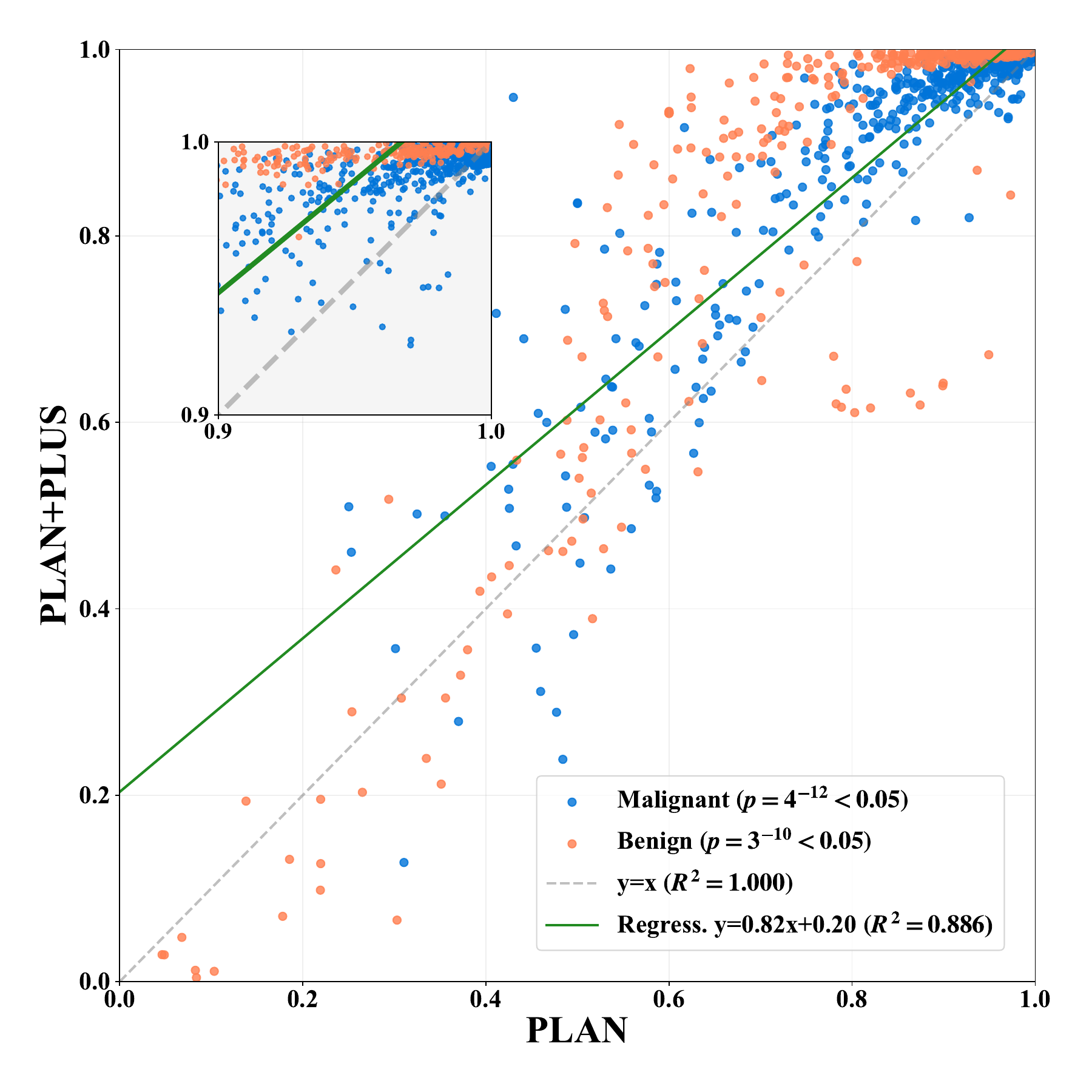}
        \caption{}
    \end{subfigure}
    \hfill
    \begin{subfigure}{0.25\textwidth}
        \centering
        \includegraphics[width=\linewidth]{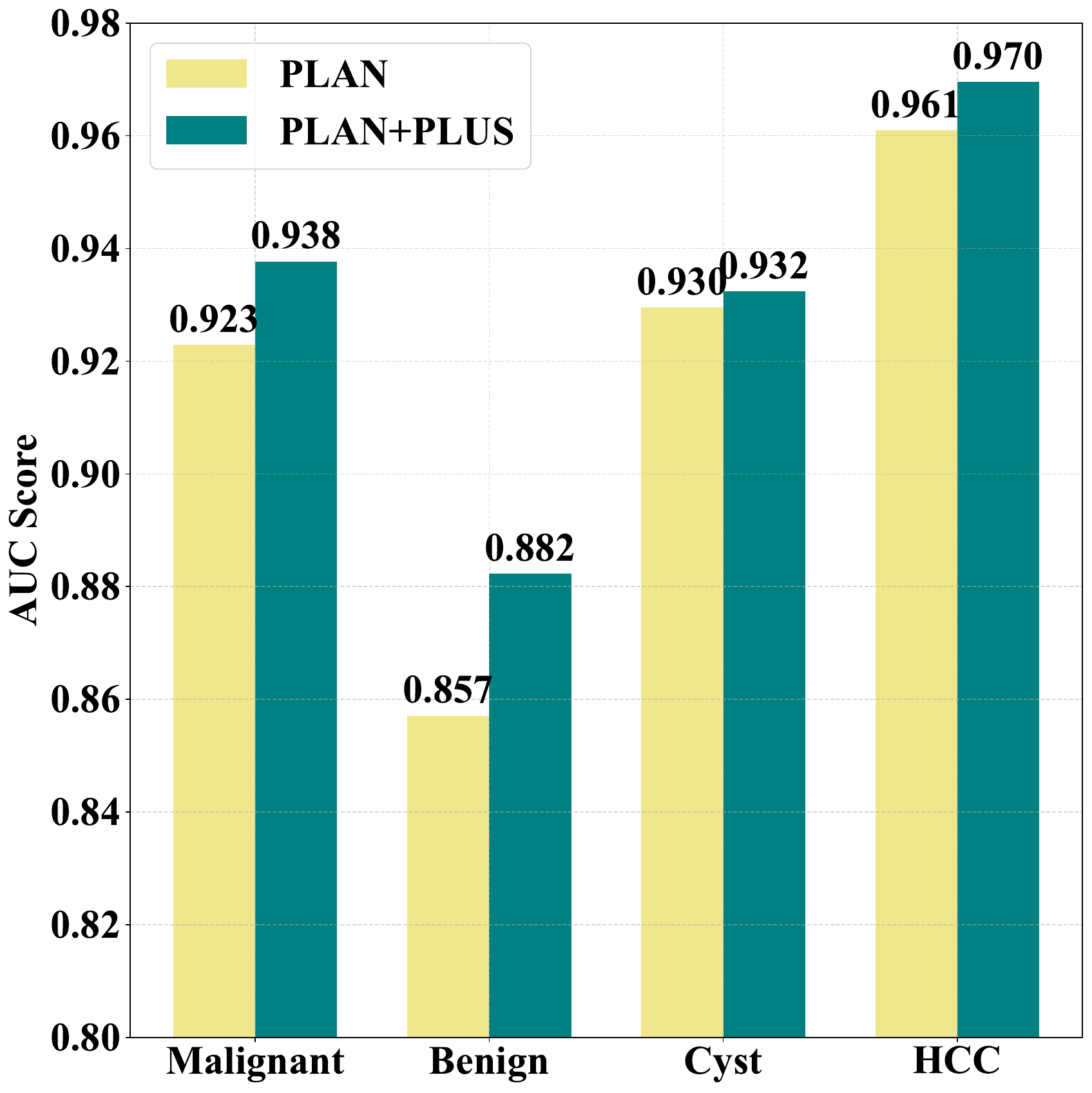}
        \caption{}
    \end{subfigure}
    \hfill
    \begin{subfigure}{0.45\textwidth}
        \centering
        \includegraphics[width=\linewidth]{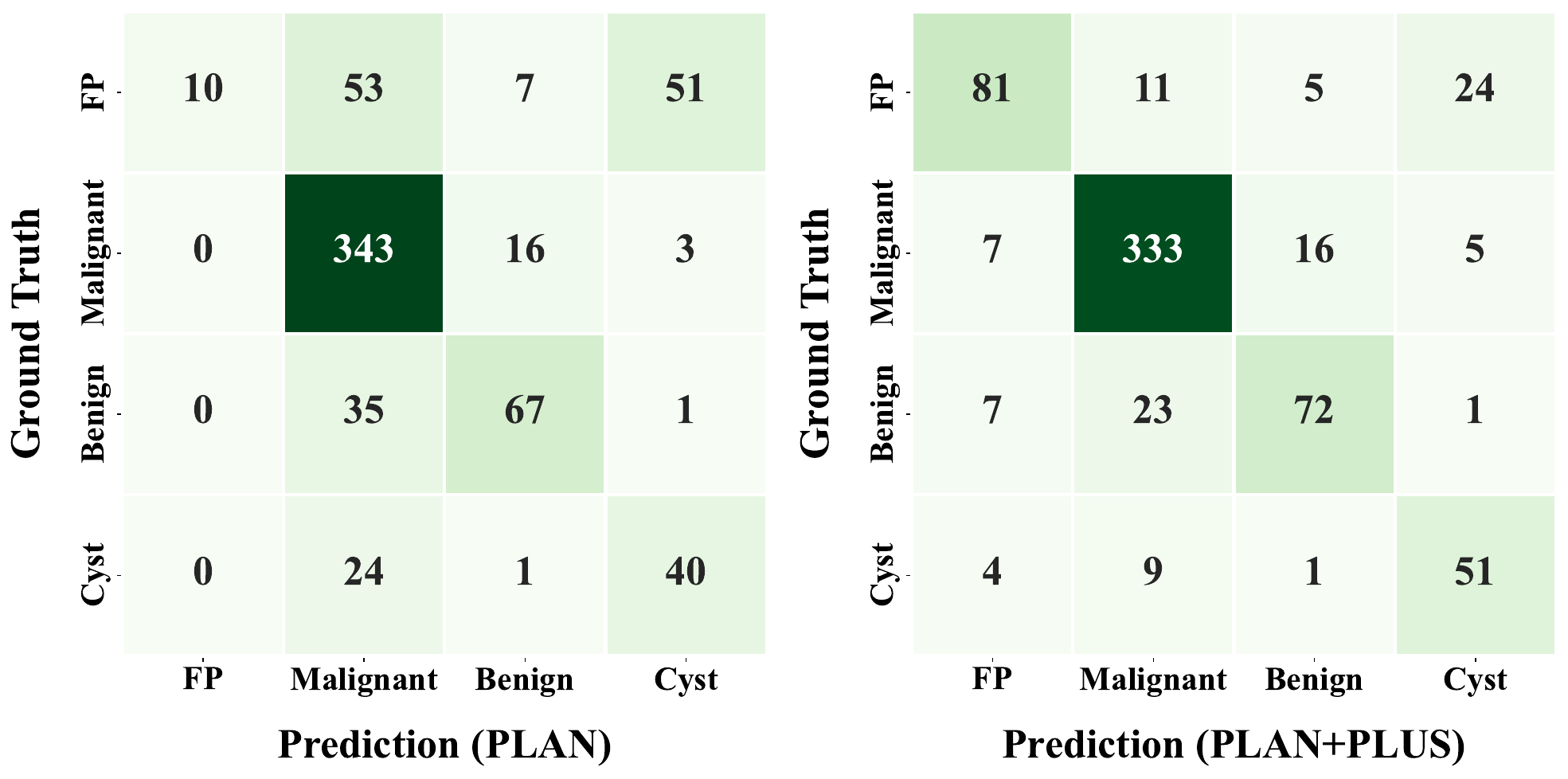}
        \caption{}
    \end{subfigure}
    \caption{Quantitative analysis. (a) Binary classification significance test. (b) Lesion-wise AUC comparisons. (c) Patient-wise confusion matrix.}
    \label{fig:quantitative}
\end{figure}

\textbf{Dataset.} We constructed a large-scale NCCT dataset comprising 8,651 patients with or without liver tumors from two medical centers\footnote{\scriptsize Ethics approval numbers by institutional review board: 2024PS954K, IIT20220356B-R2.}. To efficiently ensure high-quality annotations, we adopted a human-in-the-loop approach where two senior radiologists ($>$10 yrs experience) first manually annotated a subset of cases to train a preliminary model, which then generated results on remaining samples for iterative refinement. Ground-truth lesion diagnosis comes from pathology reports, imaging features, and follow-up data. The dataset includes 28,853 FLLs in nine types, including malignant lesions
%\footnote{\scriptsize Sarcoma, PEComa, treated tumor, adenoma, hamartoma, inflammatory hyperplasia, etc.}
: hepatocellular carcinoma (HCC; 3,321), cholangiocarcinoma (CCA;1,866), metastasis (8,299), other malignant (34); and benign lesions: hemangioma (1,381), focal nodular hyperplasia (128), calcification (755), cyst (13,049), other benign (23). We stratified the dataset into training (6,500), validation (951), and testing (1,200) sets.

\noindent
\textbf{Implementation Details.} 
The experiments were conducted on a single NVIDIA A800 GPU. All pre-segmentation models $\phi(\cdot)$ were trained using 5-fold cross-validation on the training set for 1,000 epochs. The network architecture consists of two 3D ResNet-18~\cite{He2015DeepRL} encoders, i.e, $E_\theta^S$ and $E_\theta^H$, to capture lesion-specific and global context liver features. For each detected lesion from $\phi(\cdot)$, we extracted volumetric ROIs of size $64\times64\times16$ voxels centered at the lesion. Input volumes were pre-processed using the nnUNet~\cite{Isensee2021nnu} pipeline, including resampling to isotropic voxel spacing and normalization with an abdominal window setting. The network was trained for 100 epochs using the AdamW optimizer~\cite{loshchilov2017decoupled} (learning rate=$10^{-4}$, weight decay=0.05, batch size=2) with cosine annealing scheduling. The loss weights are $\alpha=0.5$ and $\beta=0.3$. For HDA, we set $K=4$.

% \subsection{Experimental results}
\noindent
\textbf{Main Experimental Results.} 
We evaluate the proposed PLUS framework by integrating it with three off-the-shelf liver lesion segmentation baselines with high FLL detection sensitivity: nnUNet~\cite{Isensee2021nnu} is widely recognized as the de facto gold standard in medical image segmentation with its robust generalizability~\cite{isensee2024nnu}; Mask2Former~\cite{Cheng2022Mask2Former} demonstrates superior performance in instance segmentation with dense cross-attention architecture; and PLAN~\cite{Yan2023Plan} extends Mask2Former with improved anchor queries and foreground-enhanced sampling loss, achieving state-of-the-art performance in joint liver tumor segmentation and diagnosis. 

As summarized in Table~\ref{tab:performance}, PLAN+PLUS achieves the best performance across most metrics, with F1 scores of 65.11\%, 90.12\%, and 73.10\% for lesion-wise detection, malignant patient-level, and benign patient-level assessment, respectively. Notably, the plugin PLUS demonstrates consistent performance improvements when integrated with existing architectures. Compared to vanilla PLAN, PLAN+PLUS shows significant gains of 5.66\% in lesion-wise F1 score, 6.26\% in malignant patient-level F1 score, and 4.03\% in benign patient-level F1 score. Similar improvement patterns are also observed when applying the plugin module to nnUNet and Mask2Former, validating the effectiveness and architecture-agnostic robustness of PLUS.

Fig.~\ref{fig:quantitative} presents a comprehensive quantitative comparison between PLAN and PLAN+PLUS. The probability correlation analysis in Fig.~\ref{fig:quantitative}\textcolor{red}{(a)} demonstrates that PLAN+PLUS %generates significantly more reliable predictions 
increases the correct confidence score of each class than the baseline PLAN, particularly for cases in the mid-probability range where classification is typically more challenging. Fig.~\ref{fig:quantitative}\textcolor{red}{(b)} reveals substantial improvements in both malignant vs.~benign classification and the diagnosis of common lesion types. The confusion matrices in Fig.~\ref{fig:quantitative}\textcolor{red}{(c)} highlight PLAN+PLUS's superior performance through reduced false positives for malignant lesions, improved benign lesion classification, and better identification of cystic lesions, suggesting that PLUS effectively further enhances patient-wise classification accuracy.
\begin{figure}[ht!]
	\centering
\includegraphics[width=1\textwidth]{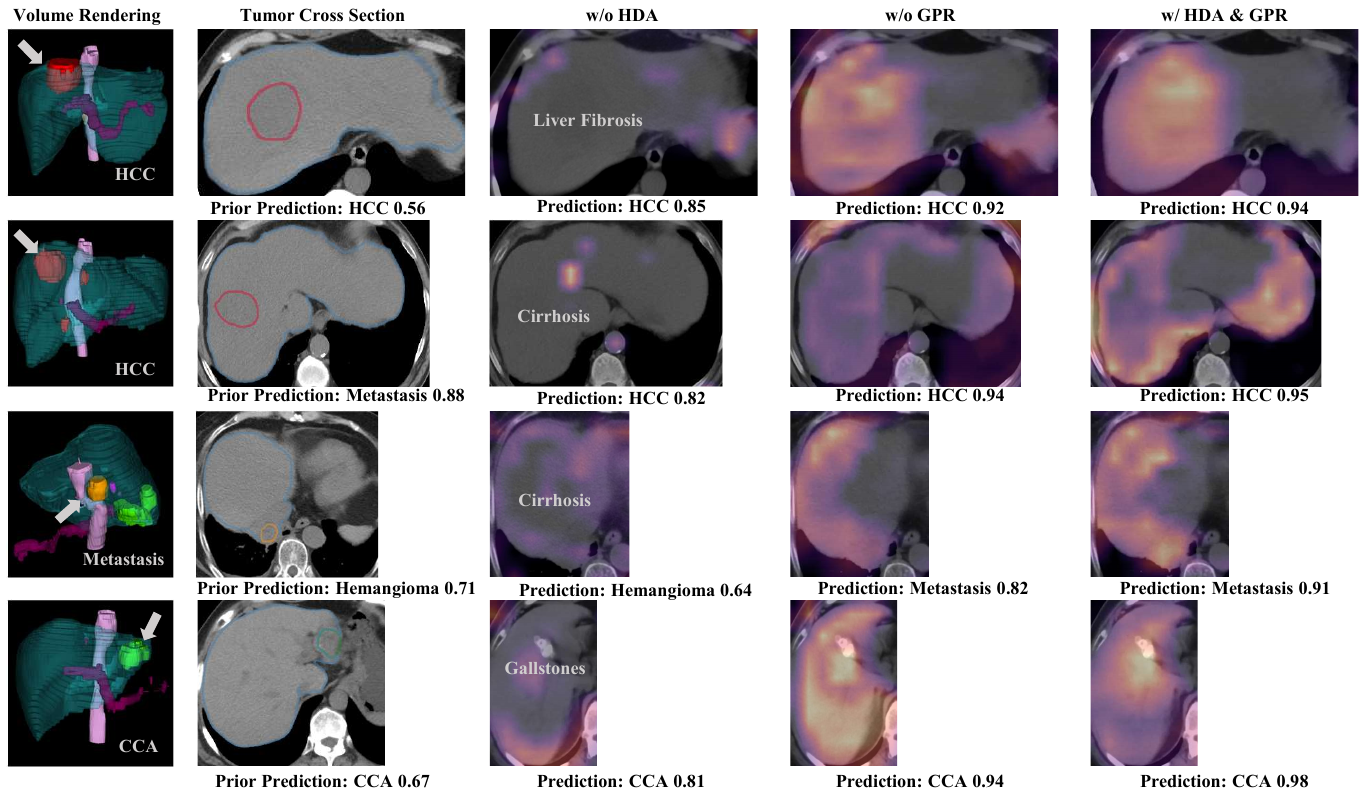}
	\caption{Qualitative results showing ablation study of model components. Each case is annotated with the predicted lesion category and its probability.}
	\label{fig:vis}
\end{figure}

\noindent\textbf{Effect of GPR.} To validate the effectiveness of GPR, we compare it with several alternative fusion strategies as shown in Fig.~\ref{fig:GPR}, including distillation~\cite{gou2021knowledge}, gated-fusion~\cite{arevalo2017gate}, and weighted fusion methods~\cite{yin2018tensor}. Given the potential inaccuracies in prior $P$ and its dimensional imbalance with $\textbf{F}$, existing approaches show limitations: knowledge distillation restricts priors to supervisory signals, gated fusion lacks adaptability to varying prototype relevance, and weighted fusion fails to capture complex feature-prior relationships. As shown in Table~\ref{tab:performance}, GPR (last row) addresses these challenges and outperforms them across different metrics through GNN-based relationship modeling and adaptive feature aggregation.

\noindent
\textbf{Ablation Study.} 
We conducted ablation studies to validate each component of the PLUS framework. Starting with the PLAN baseline, adding HDA significantly enhances lesion-level F1 by 4.40\% and malignant patient-level F1 by 3.53\%. GPR further improves performance, and after adding combined loss, despite causing slight deterioration at the lesion level, it achieves optimal results at the patient level with substantial gains in malignant and benign F1.

Saliency method visualizations (e.g., Grad-CAM~\cite{selvaraju2017grad,zhang2023revisiting}) show the baseline exhibits diffuse attention, often missing critical regions. HDA focuses on diagnostically pivotal areas (e.g., fibrotic patterns, uneven liver border that indicates cirrhosis, and gallstones that correlate with CCA) and resolves ambiguities between challenging subtypes. GPR refines attention to pathology-specific regions by fusing segmentation priors with liver features. Combined, HDA+GPR produces anatomically coherent attention maps that capture both localized lesion characteristics and structural context, confirming they enhance the model's focus on clinically meaningful regions.
\section{Conclusion}
\noindent

In this work, we present PLUS, a plugin framework that enhances existing segmentation models for NCCT FLL analysis. PLUS further improves the state of the art while maintaining architectural flexibility. Future work will extend this paradigm to multi-modal imaging integration and real-world clinical trials to support clinical decision-making and intervention.

\section*{Acknowledgement}
This work was supported by the National Natural Science Foundation of China (No. 82471971);  General Program of the Liaoning Provincial Department of Education  (LJKMZ20221160);  Liaoning Province Science and Technology Joint Plan (2023JH2/101700127); the Leading Young Talent Program of Xingliao Yingcai in Liaoning Province (XLYC2203037).

\bibliographystyle{assets/plainnat}
\bibliography{paper}

% \clearpage
% \newpage
% \beginappendix
% \input{sec/appendix}

\end{document}